\documentclass[doublecol,figures]{epl2}

\usepackage{graphicx} 
\usepackage{amsmath}
\usepackage{amssymb}
\usepackage{gensymb}
\usepackage[normalem]{ulem}

\title{Lattice dynamics in the FeSi-based family of superconductors}

\author{S. Layek (equal contribution)\inst{1,2} 
\and M. F. Hansen (equal contribution)\inst{1,3} 
\and J.-B. Vaney\inst{4} 
\and P. Toulemonde\inst{1}
\and S. Tenc\'e\inst{2} 
\and P. Boullay\inst{5} 
\and A. Cano\inst{1}
\and M.-A. M\'easson\inst{1}
}
\shortauthor{S. Layek, M. F. Hansen \etal}

\institute{                    
\inst{1} CNRS, Universit\'e Grenoble Alpes, Institut N\'eel, 38042 Grenoble, France\\
\inst{2} Department of Physics, School of Advanced Engineering, University of Petroleum and Energy Studies (UPES), Dehradun, Uttarakhand 248007, India\\
\inst{3} Earth and Planets Laboratory, Carnegie Institution for Science, 5241 Broad Branch Road, NW, Washington, DC 20015, USA\\
\inst{4} Univ. Bordeaux, CNRS, Bordeaux INP, ICMCB, UMR 5026, F-33600 Pessac, France\\
\inst{5} Normandie Universit\'e, ENSICAEN, UNICAEN, CNRS, CRISMAT, 14000 Caen, France
}

\abstract{
The lattice dynamics of the superconducting materials LaFeSiH and LaFeSiO$_{1-\delta}$ as well as their intermetallic precursor LaFeSi are investigated by polarized Raman spectroscopy and first-principles calculations, together with X-ray and advanced electron diffraction techniques for their structural analysis. We find that the Fe-dominated Raman-active modes reflect the chemical peculiarities of these silicides compared to their pnictide counterparts, with enhanced structural couplings between the FeSi layer and the spacer that can be related to the ionic vs covalent character of the latter. In addition, we find signatures of enhanced electron-phonon coupling for some of the Raman-active modes. Beyond that, our study reveals intriguing Fe-based Raman features as well as structural subtleties in LaFeSiH suggesting that this superconductor may formally be non-centrosymmetric.  
}

\begin{document}

\maketitle

\section{Introduction}

Iron-based superconductors provide a distinct realization of unconventional superconductivity \cite{Hosono2015,Alloul2016}
which, in some cases, can be supplemented with intriguing topological features \cite{Zhang2018,zhang2019}.
These properties motivate a constant interest in new materials of this class. 
At present, there are only four successful compounds in which iron is combined with non-toxic crystallogen elements: the germanide YFe$_2$Ge$_2$ with superconducting T$_{c}$ $\sim$ 0.1 K \cite{Zou2014, Kim2015, Chen2016} and the silicides 
LaFeSiH \cite{Bernardini2018,Bhattacharyya2020,Hansen2023}, LaFeSiF  \cite{Vaney2022}, and LaFeSiO with T$_{c}$ = 10 K \cite{Hansen2022}.
In the case of the silicides, some of their structure-property relationships challenge previous precepts such as the quasi-universal correlation between anion height and $T_c$ thus calling for further investigations \cite{Hansen2022}.

In this paper, we address the dynamical structural properties of the FeSi-based superconductors together with that of their LaFeSi precursor by means of polarized Raman spectroscopy combined with density-functional-theory (DFT) calculations and advanced diffraction techniques.
The measured phonon frequencies as well as the observed selection rules are found to be in good agreement with the calculations.
In LaFeSiH, however, we observe a Raman Fe-based feature displaying an intriguing double-peak shape. Driven by this observation, we revisit the crystal structure of this system by means of electron and X-ray diffraction. Furthermore, we elucidate the structural coupling between the FeSi layer and the La$X$ spacer in these systems from the trends of the phonon energies, which we discuss also in relation to the ionic vs covalent character of the spacer as a function of its precise composition ($X =$ empty site, H, or O). 
In addition, the relative strength of the electron-phonon interaction is discussed from the Fano shape observed for some of the Raman-active modes.

\section{Methods}

Single crystals of the intermetallic LaFeSi were extracted from the bulk of an arc-melted sample with nominal composition La$_{35}$Fe$_{35}$Si$_{30}$.
LaFeSiH single crystals were then obtained by hydrogenation of 
these crystals at 250°C under a gas flow of H$_2$ for 4h. 
Both the LaFeSi precursor and the LaFeSiH superconductor crystallize in the tetragonal ZrCuSiAs-type structure 
as confirmed from single-crystal X-ray diffraction.
The samples have sub-millimeter size and a plate-like shape with the larger faces perpendicular to the crystallographic \textit{c}-axis.
The LaFeSiO samples, 
however, 
are from the same batch 
as Ref.~\cite{Hansen2022}, which displays an oxygen-site occupancy of $\sim$~0.9 and a T$_c$ of 10~K. Specifically, they were obtained by heating powders of LaFeSi at 320~\textdegree{}C in 20\%/80\% O$_{2}$/Ar flow for 72~h. 
Microscopic crystals were then selected
and the shiny faces were assumed to be parallel to the (001) plane as is the case for the other compounds. 
From these crystals we obtained well-defined polarized Raman spectra. 
However, the perfect orientation of the $a$-axis is lost in this case and hence the selection rules are almost fully (but not completely) obeyed. 

The Raman spectra of small single crystals were measured under a microscope attached to a WITec (Model: alpha 300R) confocal Raman spectrometer. A 532~nm laser line was used for the excitation. The laser power was kept below 1 mW in order to avoid local overheating and/or damaging the crystals. Macro-Raman as a function of temperature was performed on a triple-stage spectrometer and in a cryo-free cryostat down to 9~K. Different polarizations of the light and orientations of the samples were investigated to access the symmetries of the excitations.

Precession electron diffraction tomography (PEDT) was performed with a JEOL F200 cold-FEG transmission electron microscope operated at 200 kV, equipped with a NANOMEGAS DigiStar precession module and a GATAN RIO16 camera. 
Samples for PEDT investigations were prepared by smoothly crushing powder. PEDT data was processed using the programs PETS 2.0 \cite{Palatinus2019} and Jana2020 \cite{Petricek2014} (Cf. Suppl. Mat.). 
 
The DFT calculations were performed with the Quantum {\sc ESPRESSO} package \cite{Giannozzi2009} using the norm-conserving pseudopotentials from the PseudoDojo library \cite{VanSetten2018}. We used the Perdew-Burke-Ernzerhof form of the generalized gradient approximation \cite{Perdew1996}. The calculations were converged with a Monkhorst-Pack mesh of 13$\times$13$\times$7 $k$-points and a 125~Ry cutoff for the wave functions with a 0.01~Ry smearing. We used the experimental lattice parameters reported in ref. \cite{Welter1992,Bernardini2018,Hansen2022} and optimized the internal coordinates of the La and Si atoms.

\section{Preliminaries} We first recall the nature of the $\Gamma$-point phonons expected in our systems and their Raman selection rules. The systems under consideration represent the FeSi-based counterparts of previous 111 and 1111 Fe-based superconductors with $P4/nmm$ crystal structure ($D_{4h}$ point group) {\cite{Ivantchev2000}. 
The contribution of the different atoms to the different $\Gamma$-point phonons is summarized in Table~\ref{tab2}.   
In the case of the LaFeSi precursor, we then have two A$_{1g}$, one B$_{1g}$ and three E$_g$ Raman-active modes. 
In LaFeSiH and LaFeSiO, there is an additional B$_{1g}$ Raman-active mode as well as another E$_g$ mode. 

In our experiments, the \textit{c}-axis of the samples is parallel to Poynting vector of the light. 
The E$_g$ modes, which are strongly affected by magnetism in BaFe$_{2}$As$_{2}$ \cite{Chauviere2009}, are 
not visible for this configuration. 
As a result, we then probe a total of 2A$_{1g}$ + B$_{1g}$
and 2A$_{1g}$ + 2B$_{1g}$ Raman-active modes in LaFeSi and LaFeSiH/O, respectively, with selection rules as summarized in Table~\ref{table_1}.

These modes are associated with out-of-plane displacements of the atoms. In the following, we denote these modes as $A_{1g}^{\rm La/Si}$ and $B_{1g}^{\rm Fe/H/O}$ according to the main contribution to these displacements. However, these modes generally have a mixed character with dominant and subdominant displacements of different atoms (Fe-$X$ and Si-La). 

\begin{table}[t!]
\caption{Atoms and $\Gamma$-point phonons to which they contribute in LaFeSi and LaFeSiH/O. The Raman-active modes are indicated in bold.}
\label{tab2}
\begin{center}
\begin{tabular}{ c c c } \hline \hline
Atom (Wyckoff position) & \textbf{${\Gamma}$}-point modes \\ \hline 
La  ($2c$) & \textbf{A$_{1g}$}+A$_{2u}$+\textbf{E$_g$}+E$_u$ \\ 
Fe  ($2b$) & \textbf{B$_{1g}$}+A$_{2u}$+\textbf{E$_g$}+E$_u$ \\ 
Si  ($2c$) & \textbf{A$_{1g}$}+A$_{2u}$+\textbf{E$_g$}+E$_u$ \\ 
H/O  ($2a$) & \textbf{B$_{1g}$}+A$_{2u}$+\textbf{E$_g$}+E$_u$ \\ \hline \hline
\end{tabular}
\end{center}
\end{table}

\begin{table}[t!]
\caption{Raman selection rules for the $D_{4h}$ point group in Porto's notation. 
$a'$ stands for the (110) axis.
}
\label{table_1}
\begin{center}
\begin{tabular}{c c c} \hline \hline
 Polarization geometry && Raman-active phonons \\ \hline 
$c(aa)\bar{{c}}$ && A$_{1g}$+B$_{1g}$\\  
$c(a'a')\bar{{c}}$ && A$_{1g}$+B$_{2g}$ \\
$c(ba)\bar{{c}}$ && A$_{2g}$+B$_{2g}$ \\ 
$c(b'a')\bar{{c}}$ && A$_{2g}$+B$_{1g}$ \\\hline \hline
\end{tabular}
\end{center}
\end{table}

\begin{figure*}[t!]
\onefigure[width=\textwidth]{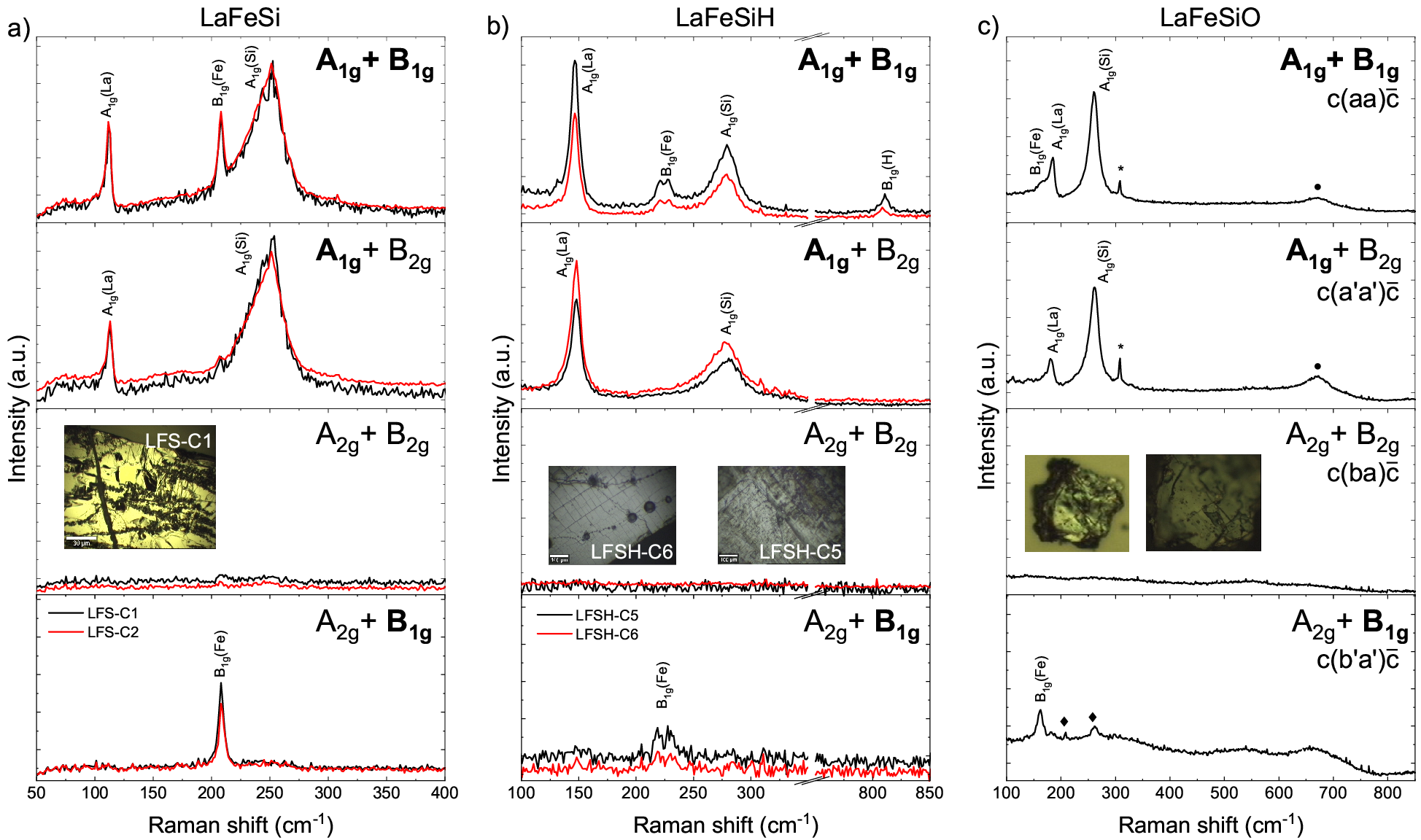}
\caption{
Polarized Raman spectra of LaFeSi, LaFeSiH and LaFeSiO single crystals measured at room temperature.
The configuration of the polarization is indicated in Porto's notation, with the selected Raman modes in bold.
Black and red curves represent data from different samples. 
The insets show optical images of the samples. 
For LaFeSiO, star and losange symbols indicate a signal from the optical fiber in the setup and `leakage' from A$_{1g}$ modes, respectively, while circles indicate an additional high-energy feature, probably related to a multi-phonon process.
}
\label{Fig1}
\end{figure*}

\section{LaFeSi precursor}

The Figure \ref{Fig1}a) shows the polarized Raman spectra obtained on single crystals of the LaFeSi precursor at room temperature. 
The symmetry of the Raman-active phonons behind the observed peaks are deduced by changing the polarization geometry according to Table~\ref{table_1}. 
We clearly identify the three modes $A_{1g}^{\rm La}$, $A_{1g}^{\rm Si}$, and $B_{1g}^{\rm Fe}$.
This identification is additionally supported by the DFT calculations (see Table \ref{t:summary}), which also provide the main atomic character of these modes.

\section{Superconducting LaFeSiH} 

The polarized Raman spectra obtained from single crystals of LaFeSiH are shown in Figure \ref{Fig1}b). 
In addition to the
three modes $A_{1g}^{\rm La}$, $A_{1g}^{\rm Si}$, and $B_{1g}^{\rm Fe}$, we observe the mode corresponding to $B_{1g}^{\rm H}$.
The frequencies of these modes are in reasonable agreement with the calculations (see Table \ref{t:summary}), which again provide the main atomic character of these modes.

\begin{table*}[t!]
\caption{Frequencies of the different $\Gamma$-point phonons of the investigated silicides together with their symmetries. 
In the case of the Raman-active modes observed in our experimental setup (i.e. $A_{1g}$ and $B_{1g}$), the main atomic displacements are indicated with the superscripts.
\label{t:summary}}
\begin{center}
\resizebox{!}{3.55cm}{
\renewcommand{\arraystretch}{1.1}
\begin{tabular}{cccc|cccc|ccccc} 
\hline \hline
 \multicolumn{4}{c|}{LaFeSi} & \multicolumn{4}{c|}{LaFeSiH} & \multicolumn{4}{c}{LaFeSiO} \\ 
 \multicolumn{2}{c}{Frequency (cm$^{-1}$)} && Mode & \multicolumn{2}{c}{Frequency (cm$^{-1}$)} && Mode & \multicolumn{2}{c}{Frequency (cm$^{-1}$)} && Mode \\ 
Experimental & Calculated && symmetry & Experimental & Calculated && symmetry & Exp. & Cal. && symmetry  \\ \hline
       & 96 
       && $E_g$                &     & 98 
       && $E_g$             &   & 79 && $E_u$ \\
 112.6 & 123 
 && $A_{1g}^{\rm La}$ &      & 115 
 && $E_u$             &   & 106 && $A_{2u}$\\
       & 148 
       && $E_{u}$           & 146.4 & 120 
       && $A_{1g}^{\rm La}$&   & 123 && $E_g$ \\   
       & 152 
       && $A_{2u}$             &      & 122 
       && $A_{2u}$ &164.4 & 156 && $B_{1g}^{\rm Fe}$  \\ 
       & 209 
       && $E_g$              &    & 178 
       && $E_g$             &  & 179 && $E_{g}$ \\
205.6 & 218 
&& $B_{1g}^{\rm Fe}$  & 222.8, 228.5 & 228 
&& $B_{1g}^{\rm Fe}$&186 & 186 && $A_{1g}^{\rm La}$ \\   
250.0 & 313 
&& $A_{1g}^{\rm Si}$ & 279   & 289 
&& $A_{1g}^{\rm Si}$& -- & 272 && $B_{1g}^{\rm O}$ \\
    & 322 
    && $A_{2u}$             &    & 315 
    && $A_{2u}$           &    & 282 && $A_{2u}$ \\
    & 364 
    && $E_u$               &    & 403 
    && $E_u$               &    & 295 && $E_u$\\
    & 372 
    && $E_g$               &    & 407 
    && $E_{g}$             &260 & 297 && $A_{1g}^{\rm Si}$\\
    &           &&                     & 809.2 & 792 
    && $B_{1g}^{\rm H}$ &    & 389 && $A_{2u}$\\
    &           &&                     &       & 809 
    &&  $E_u$           &    & 414 && $E_g$ \\
    &           &&                     &       & 822 
    && $A_{2u}$         &    & 414 && $E_u$\\
    &           &&                     &       & 847 
    &&  $E_g$           &    & 433 && $E_{g}$\\
    \hline \hline 
\end{tabular}
}
\end{center}
\end{table*}
\begin{figure}
\onefigure[width=8cm]{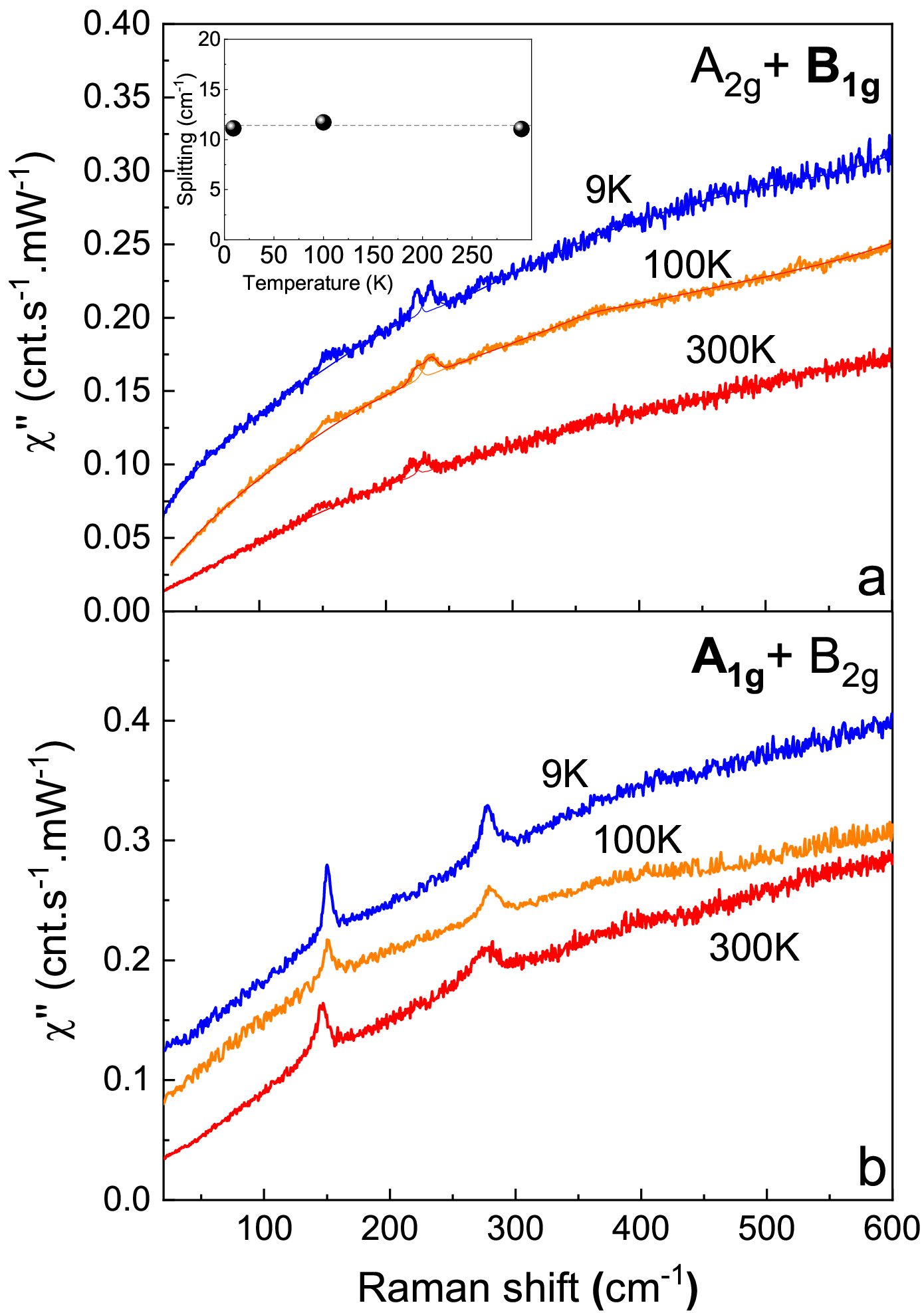}    
\caption{Temperature dependence of the polarized Raman spectra on a third LaFeSiH single crystal, at 300~K, 100~K and 9~K. Spectra are shifted for clarity. 
The inset in a) shows that the 
distance between the maxima of the two-peak feature at 228~cm$^{-1}$ is constant with temperature. 
} 
\label{Fig3}
\end{figure}

In the measured spectra, however, the $B_{1g}^{\rm Fe}$ mode at $\sim$228~cm$^{-1}$ appears as a double-peak feature. 
This is surprising because, even if this mode involves dominant Fe and subdominant H displacements, it 
should be a single non-degenerate mode.
{Accordingly, a symmetry breaking, for example, cannot produce such splitting. 
Furthermore, as shown in Fig. \ref{Fig3},
no new modes or new splittings appear as a function of temperature and in none of the four probed symmetries, which is consistent with the absence of a structural transition associated with magnetism reported in \cite{hansen2023magnetic}. 
The B$_{1g}$ double-peak feature, however, persists down to 9~K as seen Fig.~\ref{Fig3}(a) and both the position of its two maxima and overall width remain quite constant. 
Note that the Raman inactive modes are relatively far in frequency from this feature (see Table \ref{t:summary}), so that the Raman activation of some of these modes due to a structural change could hardly explain this observation (unless the change would be quite substantial).

The double-peak at $\sim 228$~cm$^{-1}$ is robust, in the sense that it is systematically observed in all the samples at the same frequency with similar relative intensities. 
Thus, even if the light only penetrates $\sim$~40~nm in this type of compounds, it should not be a surface artifact. 
In fact, the incomplete hydrogenation at the surface as well as a partial oxydation can be excluded by noting that the frequencies and FWHM of corresponding B$_{1g}$ modes in the LaFeSi precursor and LaFeSiO (see below) are very different. 
Thus, we conclude that such a double-peak feature is LaFeSiH specific and has an intrinsic character.

However, according to the above, that feature cannot be easily explained within the $P4/nmm$ structural model of LaFeSiH and hence questions the accuracy of such a model. 
Thus, we used PEDT to carefully revisit the structure of this system. As explained in the Supplemental Material,
the analysis of the PEDT data confirms the atomic positions and the site occupancies previously derived from X-ray diffraction \cite{Bernardini2018}.
However, it also reveals the presence of additional weak reflections as illustrated in Fig. \ref{FigPEDT}, that suggests the absence of the $n$-glide plane symmetry.
The PEDT refinement also displays a residual density close to the Si sites, if the $P4/nmm$ structure is assumed.
These observations thus confirm that the ideal $P4/nmm$ model does not accurately capture all the structural details of the samples. 
To confirm the presence of these extra reflections, inferring that the actual structure would be non-centrosymmetric}, we then performed additional single-crystal X-ray diffraction measurements.

However, due to their very weak intensity in both PEDT and X-ray diffraction, we could not determine the crystal structure of these systems more accurately. At the same time, we note that the extra reflections could also be due to the presence of stacking faults along $c$.  
In any case, the actual structure of these systems seems to be immeasurably close to $P4/nmm$ while the Raman double-peak feature observed in LaFeSiH is visibly a much stronger effect. These observations then remain open questions.
\\

\begin{figure}[t!]
\onefigure[width=.475\textwidth]{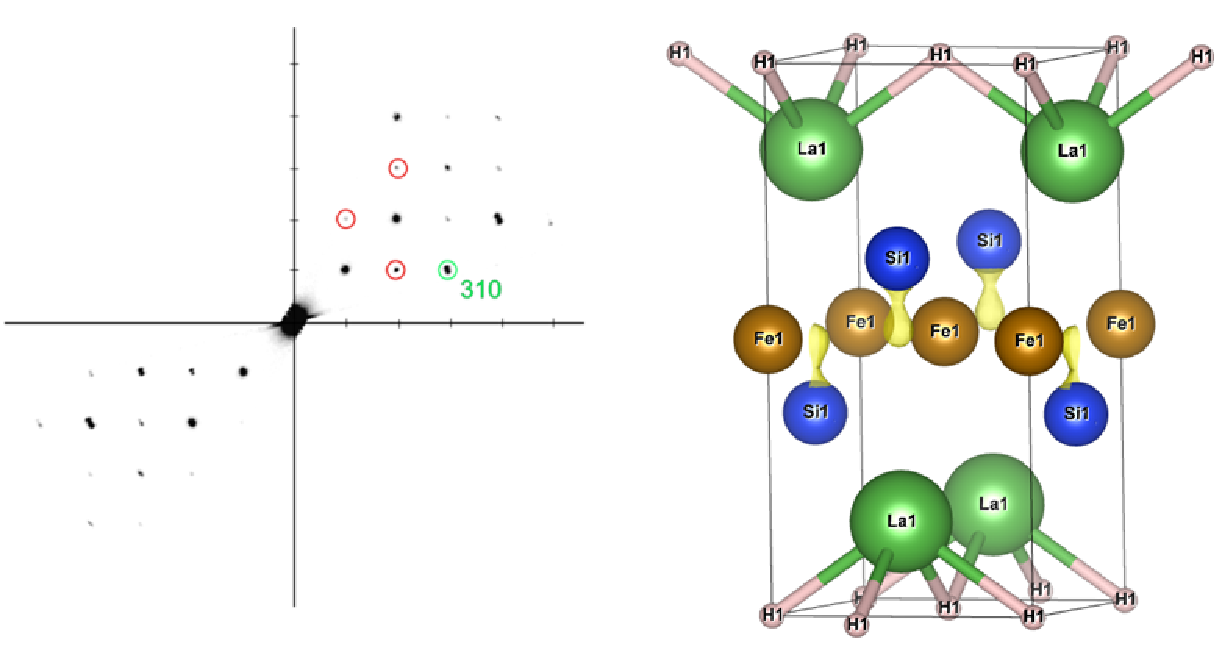}
\caption{
Electron diffraction pattern obtained for LaFeSiH at the $hk0$ plane. The green circle indicates the 310 reflection expected according to the $P4/nmm$ structure. 
The reflections encircled in red, however, are incompatible with $n$-glide plane symmetry. 
The Fourier-difference map obtained after the PEDT refinement is superimposed to the ball-and-stick model of the corresponding $P4/nmm$ structure using VESTA\cite{Momma2011}. This map shows residual densities close to Si sites (isosurface levels with 3$\sigma$[$\Delta$V(r)] are in yellow).  
} 
\label{FigPEDT}
\end{figure}

\section{Superconducting LaFeSiO} 

The polarized Raman spectra obtained from LaFeSiO are shown in Fig. \ref{Fig1}c). First, in the parallel polarized configuration we clearly observe two A$_{1g}$ modes, at 187~cm$^{-1}$ and 260~cm$^{-1}$. The mode measured at 165~cm$^{-1}$ is assigned to the 
$B_{1g}^{\rm Fe}$ mode. 
In these measurements, however, the polarization of light with respect to the $a$ axis is not perfect, so that there is no full extinction of the B$_{1g}$ modes. Yet, we observe 
significant changes in the peak intensities from which the symmetry of the corresponding modes can be deduced. The relative intensity of the A$_{1g}$ modes, for example, remains constant in parallel polarization when the polarization angle changes within the $ab$ plane. 

The $B_{1g}^{\rm O}$ mode is not immediately identifiable given the present data. This mode is expected at 272~cm$^{-1}$ according to our calculations and should be observable in the same configuration as the 
$B_{1g}^{\rm Fe}$ mode. We do observe intensity at 262 cm$^{-1}$ but it is more likely attributed to a "leakage" of the 
$A_{1g}^{\rm Si}$ (marked with a losange in Fig.~\ref{Fig1}), similarly to the "leakage" of the 
$A_{1g}^{\rm La}$ 
mode
measured at about 182 cm$^{-1}$. 
In addition, a broad peak, marked with a circle, is measured at 672~cm$^{-1}$. It seems to be active in the A$_{1g}$ channel. 
When considering leakage of E$_g$ modes, the closest E$_g$ mode is calculated to be at 433 cm$^{-1}$ and hence excludes such a possibility. In fact, there is no $\Gamma$-point phonon matching that frequency.
This high-energy feature may thus be related with a scattering process involving multiple phonons with finite wavevectors.

\section{Comparative analysis}
\label{disc}

To gain further insight about the structural properties of the above systems, we perform a comparative analysis in this section. The most relevant Raman features are summarized in  Fig. \ref{Fig5} (see also Table \ref{t:summary}).

First, we discuss the modes associated with the FeSi layer. 
We note that the frequency of the 
$B_{1g}^{\rm Fe}$ mode 
in LaFeSi and LaFeSiH is comparable to that of the reference compounds LaFeAsO (201~cm$^{-1}$) and SmFeAsO (208~cm$^{-1}$) reported in \cite{Hadjiev2008}. This mode, however, becomes considerably softer in LaFeSiO. This dynamical peculiarity adds to another, static one: a strikingly} reduced distance between the Si and the Fe plane in this system \cite{Hansen2022}. 
Further, we note that the $B_{1g}^{\rm Fe}$ and $A_{1g}^{\rm As}$ modes are quasi degenerate in the above arsenides, while the frequencies of the $B_{1g}^{\rm Fe}$ and the $A_{1g}^{\rm Si}$ modes are quite different in the silicides with the latter being harder (also compared with the $A_{1g}^{\rm As}$ ones).
The non-degeneracy of these modes suggests that the structural coupling of the FeSi layer to the spacer is more important in the silicides. The above trends are confirmed in the calculations.

Next, we discuss the modes associated with the spacer. The $A_{1g}^{\rm La}$ 
becomes progressively harder from LaFeSi, to LaFeSiH then to LaFeSiO. 
This can be related to the insertion of the light element within the spacer and its electronegativity. The observed trend then suggests that the ionic character of the spacer increases with increasing electronegativity of the inserted element, thereby hardening the 
$A_{1g}^{\rm La}$ mode.  
At the same time, we also note that the frequency of this mode in LaFeSiO is higher compared to that in LaFeAsO. 
This can be associated with the subdominant contribution of the Si displacements, which again suggests an enhanced structural coupling between the FeSi layer and the spacer. Since this is eventually determined by the corresponding chemical bonds, 
the electronic structure of these silicides can be expected to be more three-dimensional in general with low-energy details near the Fermi level more sensitive to changes in the spacer.

\begin{figure}[t!]
\onefigure[width=8cm]{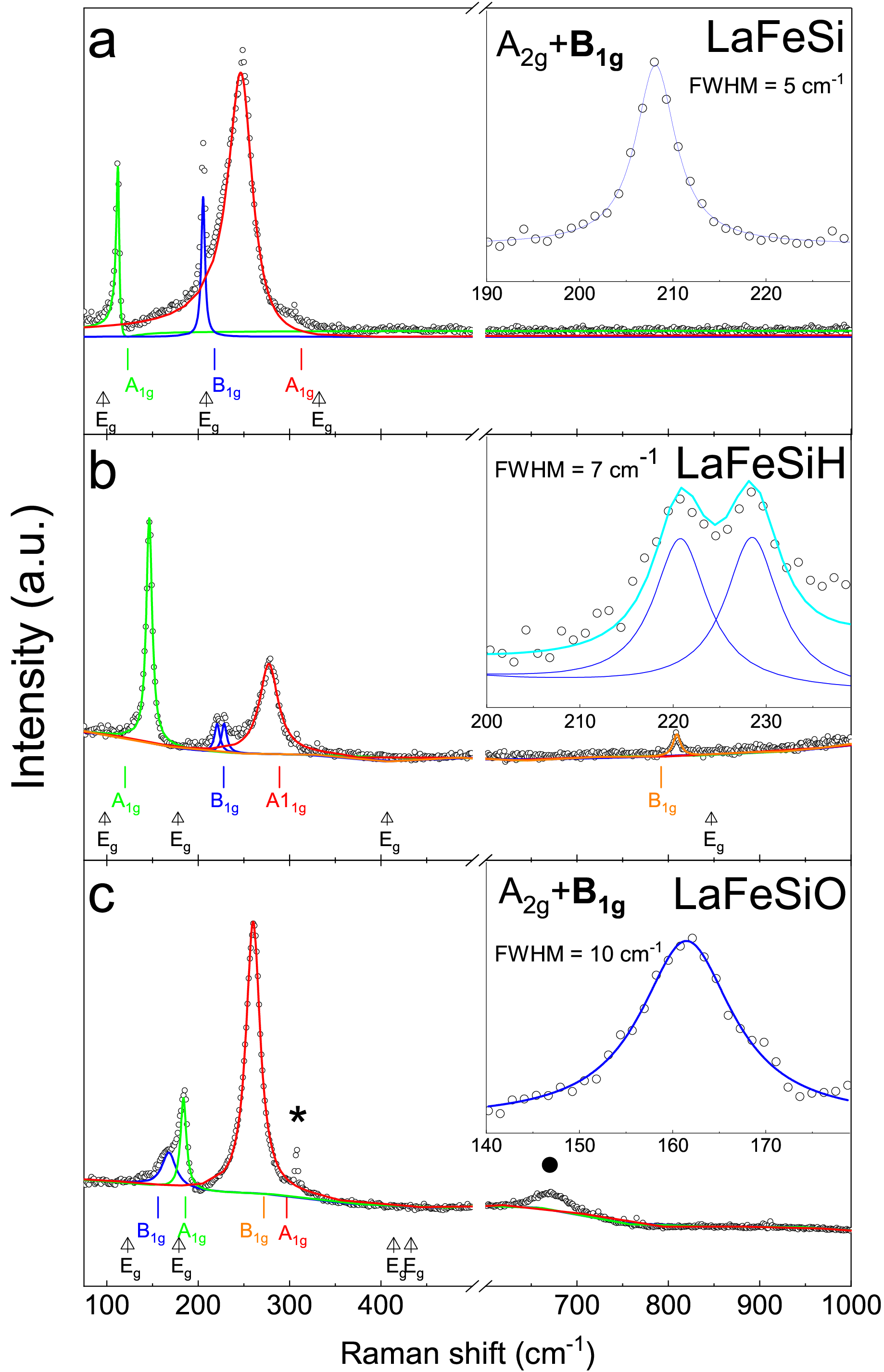}
\caption{Experimental Raman spectra of (a) LaFeSi, (b) LaFeSiH and (c) LaFeSiO 
as measured at T = 300~K. The theoretically calculated positions of the phonon modes are marked below the data as ticks and arrows for the non-degenerate and degenerate modes, respectively. The insets show the Raman features associated with the B$_{1g}^{\rm Fe}$ mode, whose fitting requires two peaks in LaFeSiH.}
\label{Fig5}
\end{figure}

\begin{table}[t!]
\caption{Fano parameter $10^3\cdot\frac{1}{q^2}$ for the two A$_{1g}$ modes mainly associated with La and Si. The corresponding fits are shown in Fig.~3 of the Supplementary Material. The higher $\frac{1}{q^2}$, the stronger the Fano effect is.}
\label{table_3}
\begin{center}
\begin{tabular}{l cc}
\hline \hline 
& A$_{1g}^{\rm La}$ & A$_{1g}^{\rm Si}$ \\ 
\hline
LaFeSi   & 32(3.4)   & 11(2.1)   \\
LaFeSiH  & 3.5(0.8)  & 3.5(0.4)     \\
LaFeSiO  & 27(4.4)   & 2.1(0.4) \\   
\hline \hline 
\end{tabular}
\end{center}
\end{table}
 
Finally, we note that, in general, the $A_{1g}$ Raman peaks in the silicides display a Fano shape that is more apparent compared to their arsenide counterparts \cite{Hadjiev2008}
(see Table~\ref{table_3}).
In addition, we observe a significant broadening of the $A_{1g}^{\rm La}$ mode from 5~cm$^{-1}$ in the LaFeSi precursor to $\sim$10.5~cm$^{-1}$ in LaFeSiH and LaFeSiO. 

These differences can be due to enhanced electron-phonon couplings for these particular $\Gamma$-point modes. 
In relation to superconductivity, however,
the most relevant couplings involve the Fe-dominated modes and, in any case, the strength of the overall coupling according to DFT calculations remains too weak to explain the superconducting $T_c$ of LaFeSiH for example \cite{PhysRevB.97.224501}.

\section{Conclusion}

Our combined analysis of the lattice dynamics of Fe-based superconducting silicides reveals that, compared with their arsenide counterparts, the structural coupling between the FeSi layer and the spacer is stronger. 
This dynamics is also found to be affected by the change from covalent to more ionic bonds within the spacer across the LaFeSi$X$ series ($X =$ empty site, H, and O). Consequently, despite their layered structure, the effective 2D behavior is different in these systems.  
Intriguingly, the Raman spectrum of LaFeSiH displays a double-peak feature that cannot be easily explained according to the ideal $P4/nmm$ crystal structure of these systems. 
Besides, detailed electron and X-ray diffraction suggests that the structure may in reality be non-centrosymmetric. 
Thus, it would be interesting to determine the structure of these systems more accurately and, if the absence of centrosymmetry is confirmed, to further investigate the superconducting properties of Fe-based silicides from that perspective.

\acknowledgments
This work was supported by the French Agence Nationale de la Recherche (ANR IRONMAN, Grant No. ANR-18-CE30-0018-03) and the Labex LANEF du Programme d’Investissements d’Avenir (grant No. ANR-10-LABX-51-01). We thanks Etienne Gaudin for fruitful discussion.

Contributions: MAM conceived the work and supervised the project with AC. JBV and ST synthetized the LaFeSi(H) samples and performed XRD measurements. MH and PT synthetized LaFeSiO. PB performed and analysized the PEDT measurements. SL  and MH performed the Raman experiments with the support of MAM. AC performed the ab-initio calculations. MAM, AC and MH discussed the results and wrote the manuscript, with contributions from all the authors.

\bibliographystyle{eplbib.bst}

\end{document}